# The role of critical current on point contact Andreev Reflection spectrum between a normal metal and a superconductor

G. Sheet, S. Mukhopadhyay, P. Raychaudhuri[a]
*Department of Condensed Matter Physics and Materials Science*
*Tata Institute of Fundamental Research*
*Homi Bhabha Rd., Colaba, Mumbai 400005, India.*

*Abstract*

The point contact spectrum between a normal metal and a superconductor often shows unexpected sharp dips in the conductance at voltage values larger than the superconducting energy gap. These dips are not predicted in the Blonder-Tinkham-Klapwizk (BTK) theory, commonly used to analyse these contacts. We present here a systematic study of these dips in a variety of contacts between different combinations of a superconductor and a normal metal. From the correlation between the characteristics of these dips with the contact area, we can surmise that such dips are caused by the contact not being in the ballistic limit. An analysis of the possible errors introduced while analysing such a spectrum with the standard BTK model is also presented.

[a]electronic mail:pratap@tifr.res.in

Andreev reflection is a process by which an electron incident from a normal metal on a normal metal/superconductor interface with energy less than the superconducting energy gap ($\Delta$) gets reflected back as a hole with opposite spin, while creating a Cooper pair inside the superconductor. Measurement of Andreev reflections using a point contact between a normal metal and a superconductor has long been used as a probe for conventional and unconventional superconductors[1-7]. In these kind of measurements, a fine tip made up of a normal metal (superconductor) is brought in mechanical contact with a superconductor (normal metal) and the differential conductance (G=dI/dV) versus voltage (G-V) characteristic of the microcontact is analysed to obtain useful informations regarding the superconductor, such as the value of the superconducting energy gap, symmetry of the order parameter etc. Recently it has been shown that this technique can also be used to obtain information on the spin polarisation of a ferromagnet[8,9] by measuring the G-V characteristic of a ferromagnet/s-wave superconductor point contact. Point contact Andreev reflection (PCAR) technique has been put to effective use to explore novel superconductors such as $MgB_2$ and superconducting borocarbides[3], heavy fermions[4,5] as well as to measure the spin polarisation in half metallic ferromagnets like $CrO_2$ [10] and $La_{0.7}Sr_{0.3}MnO_3$ [11].

PCAR G-V spectrum between a normal metal and an s-wave superconductor is usually analysed in the framework of the Blonder-Tinkham-Klapzwik[1] (BTK) theory which assumes that an electron does not undergo any inelastic scattering within a spherical volume of the diameter (i.e. *a*) of a given point contact. This can be achieved when the contact is in the ballistic limit, i.e. when the diameter (*a*) of the point contact is smaller than the electronic mean free path (*l*) in the solid. The BTK theory predicts that for a clean contact between a normal metal and a s-wave superconductor, the conductance for voltages below the superconducting gap (V < $\Delta/e$) gets enhanced by a factor of two over that in the normal state (V >> $\Delta/e$) due to Andreev reflection. For a real contact, a potential barrier almost always exists between the two electrodes originating from both an oxide barrier at the interface as well as from the Fermi wave vector mismatch between the normal metal and the superconducor. This potential barrier, modelled within the BTK formalism as a delta function barrier of the form $V(x)=V_0\delta(x)$ at the interface, causes a supression of the enhancement in G(V) below the gap value, and two symmetric peaks about $V = 0$ appear in the PCAR spectrum. An experimental spectrum is normally fitted with the BTK model using the strength of the potential barrier (expressed in terms of the dimentionless quantity $Z = V_0/\hbar v_F$, where $v_F$ is the Fermi velocity in the superconductor) and $\Delta$ as fitting parameters. According to the BTK theory, for large values of this scattering barrier ($Z\rightarrow\infty$), the position of the two peaks in the conductance gives the gap value of the superconductor. For intermediate values of Z, these peaks occur at energies slightly below $\Delta$. When a ferromagnetic metal is used as the normal metal electrode, all the Andreev reflected holes cannot propagate in the normal metal due to the difference between spin up and spin down density of states at Fermi level. This causes a suppression of the differential conductance for $V < \Delta/e$. In this case, the spectrum can be fitted with a modified BTK model[12-14], where the transport spin polarisation of the ferromagnet ($P_t=(N_\uparrow v_{F\uparrow}-N_\downarrow v_{F\downarrow})/(N_\uparrow v_{F\uparrow}+N_\downarrow v_{F\downarrow})$) is used as a fitting parameter in addition to Z and $\Delta$. In either case, no structure, apart from a smooth decay of the conductance to its normal state value, should appear in the spectrum above the superconducting energy gap.

In practice, the measured PCAR G-V spectrum often shows sharp dips in conductance[2,4-8,15-19], which cannot be easily accounted for within the ambit of the BTK formalism. These dips often appear at energies larger than the superconducting energy gap and have been observed in a wide variety of combinations between normal metals and low and high $T_c$ superconductors, such as, Nb/Cu[8], Nb/Pt[19], Pt-Ir/$Bi_2Sr_2CaCu_2O_{8+\delta}$[2], Au-$MgB_2$[15], as well as in combinations of normal metal tips and heavy fermion superconductors[4,5,18]. For a contact made with a conventional s-wave superconductor, the superconducting proximity effect[16] in the normal metal and the intergrain Josephson tunneling[17] when the superconducting electrode is polycrystalline, have been proposed as possible explanations for these dips. However, a detailed satisfactory understanding of the origin of these dips is still lacking. This hinders the extraction of reliable informations on $\Delta$ or $P_t$ from a PCAR spectrum.

In the current work, we present a systematic study of the above stated dip structures in point contacts made up of conventional superconductors and ferromagnetic and non-ferromagnetic normal metals. The point contacts were made by pressing the tip on the sample using a 100 threads per inch differential screw arrangement in a liquid He cryostat in which the temperature and the magnetic field could be conveniently varied and controlled. For point contacts on superconducting samples, a mechnically cut Pt-Ir wire was used as the normal tip. For the normal samples the PCAR spectra were measured by making contacts either with electrochemically etched Nb tips or with mechanically cut Ta tips. A four probe modulation technique operating at 362 Hz was used to directly measure the differential resistance ($R_d \sim dV/dI$) versus V characteristics, from which the differential conductance (G) was calculated, e.g. $G=1/R_d$.

In Figs. 1(a) to (d), we show some typical point contact spectra between Nb/Ta tips and Au, $Au_{1-x}Fe_x$ and Fe foils. Figure 1(e) and (f) show the spectra on a superconducting $V_3Si$ single crystal and a polycrystalline $Y_2PdGe_3$[20] sample respectively taken with Pt-Ir tip. All the spectra exhibit sharp dips at voltage values above the superconducting energy gaps (as marked by arrows). The dips are qualitatively similar in all these spectra. The dips disappear close to the superconducting transition $T_c$ or $H_{c2}$ of the superconductor. The observation of sharp dips in Nb/Fe (see panel (c)) where Fe acts as a strong pair breaker, rules out the possibility of superconducting proximity effect[16] playing a significant role in the origin of these dips. Also, the observation of the dips in single crystalline $V_3Si$ sample rules out the the possibility of intergrain Josephson tunneling[17] being a primary cause of these dips.

To investigate whether these dips are caused by the point contact not being in the pure ballistic limit, we studied the G-V spectra of a Ta/Au and a Fe/Nb point contacts by successively reducing the diameter of the point contact. To obtain a series of successive spectra the superconducting tip was initially pressed on to a Au/Fe foil giving a low resistance, large area contact. The tip was then gradually withdrawn in small steps so as to reduce the contact area (i.e. increasing contact resistance) without breaking the contact, and the spectra were recorded for each successive contacts. Figure 2 (a) and (b) show the spectra obtained in this way for Ta/Au and Fe/Nb point contacts respectively. For clarity, we have plotted here $R_d$ versus V instead of the G-V plots. Though the softness of Au allowed a better control of the point contact diameter in the Au/Ta contact, a general trend is easily discernible in the two sets of spectra. For low resistance, large area contacts the two symmetric dips in the conductance (appearing as peaks in $R_d$) appear at voltage values larger than the respective superconducting energy gaps (i.e. 0.45 meV for (a) and 1.5 meV for (b)). As the point contact diameter is reduced these dips gradually disappear and the spectra tends towards the spectra predicted by BTK[21] theory.

To comprehend the gradual emergence of the dips with increasing point contact diameter, we note that a point contact between the two metals can be categorised into three broad regimes[18,22] depending on the size *a*. In the ballistic regime, where $l \gg a$, an electron can accelerate freely within a length *a* from the point contact, with no heat generated in the contact region. For two normal metals (or a metal and a superconductor at voltages $V \gg \Delta/e$) the contact resistance in this limit is given by the Sharvin resistance $R_s = 2(h/e^2)/(ak_F)^2$. In the opposite scenario, when $l \ll a$, the potential varies smoothly over a radius *a* of the point contact due to the inelastic scattering. In this case, power gets dissipated in the contact region, thereby increasing the effective temperature of the point contact. The contact resistance in such a circumstance is governed by the Maxwell resistance, $R_M = \rho(T_{eff})/2a$, where $\rho(T)$ is the bulk resistivity and $T_{eff}$ is the effective temperature of the point contact given by $T_{eff}^2 = T^2 + V^2/4L$ (where L is the Lorentz number). When the situation does not conform to one of these two extreme regimes, the contact resistance is given by $R = R_s + \Gamma(l/a)R_M$ where $\Gamma(l/a)$ is a slowly varying function of the order of unity. Since $R_s \sim (1/a)^2$, whereas $R_M \sim (1/a)$ the Sharvin contribution to the resistance will increase more rapidly than the Maxwell contribution with decreasing contact area and for very small area it will go towards the pure ballistic limit. In between these two regimes there also exists a diffusive regime, for which the contact diameter is smaller than the inelastic scattering length, but, is larger than the elastic mean free path.

In this case, no significant heating occurs at the contact, but the Andreev reflection is suppressed as compared to that in the ballistic case[12,13]. A point of caution here is that the relationship for $R_M$ strictly holds only for contacts between similar metals. For dissimilar metals, an effective ρ(T) is ought to be substituted which could be a weighted average of the ρ(T) of the two metals.

Within the above scenario, it is now possible to account for the gradual surfacing of the dips with the increase in the resistance value of the point contact. In the data of superconductor-normal metal contacts as shown in fig. 2(a) and (b) the tip is initially pressed on the sample to generate a low resistance-large area contact. These contacts are expected to be in the thermal regime, where the point contact resistance is determined by the bulk resistivity of the two electrodes. At low current values through the point contact, the resistivity of the superconductor is zero. The contact resistance will therefore have a contribution from $R_s$ and a small contribution from $R_M$ coming fom the finite resistivity of the normal electrode. However, as the transmitted current through the point contact reaches the limiting critical current value ($I_c$) of the superconductor, the resistivity of the superconducor rapidly increases to its normal state value. Therefore, as the current reaches $I_c$, one would expect a sharp rise in the voltage across the junction, and consequently a dip in the differential conductance (G=dI/dV). As the differential screw making the point contact is gradually withdrawn, the contribution of $R_M$ in the point contact resistance decreases and the contribution of the Andreev current increases. Since the $R_M/R_s$ ratio decreases with decreasing *a* the dips become smaller and the spectrum takes the shape in conformity with BTK theory. To illustrate this point further, we have simulated the differential resistance versus voltage characteristics of the point contact, assuming that above the critical current of the superconducting tip, the voltage across the point contact consists of both the Andreev reflection contribution and the Maxwell contribution from the finite resistivity of the superconductor. The I-V contribution from Andreev reflection is calculated from the BTK model. For the superconductor above the critical current, a typical I-V curve such as the one shown with solid line in Figure 3(a) is assumed. For the I-V characteristic of a contact with a particular $R_M/R_s$ ratio at high bias voltages is calculated by adding these two voltage contributions with appropriate weight factors. The differential resistance versus voltage for different values of $R_M/R_s$ at $V>>\Delta/e$ calculated by differentiating the I-V curves generated in this way is shown with open circles in Fig. 3(a). Though the assumed I-V curve of the superconductor is empirical, the trends in figure 3(a) conforms to the experimental data: With increasing contribution from $R_M$ (i) two pronounced peaks appear in $R_d$ at $V>\Delta/e$ (marked by arrows in panel (a)) and (ii) there is an increase in the relative enhancement in the zero bias conductance compared to its high bias value. It is interesting to note that if a contact is made between a good normal metal and a superconductor with very large normal state resistivity (where the contact is likely to be in the thermal regime due to the short mean free path in the superconductor and $R_M/R_s>1$), a several fold enhancement in $G(V=0)$ value compared to $G(V>>\Delta/e)$ is expected arising from critical current alone. Such a behaviour is evident in Figure 1(f) where a 5-fold enhancement is present in a contact made between a $Y_2PdGe_3$ polycrystalline sample (with normal state resistivity $\sim 400\mu\Omega-$cm) and a Pt-Ir tip.

In the above context it becomes pertinent to carefully examine the analysis of the point contact spectra in the presence of dips in the conductance. It is apparent from Figure 2(a) that even for the smallest diameter Au/Ta contact which we could stabilise, the conductance has a finite contribution from $R_M$. The G-V curve calculated from the BTK model is indeterminate within a proportionality constant, which depends on the contact diameter as well as on the density of states and Fermi velocities in the two metals. A general practice while analysing a point contact spectrum comprising of dips is to fit it with the BTK model while ignoring these dips, using $\Delta$ and Z as the fitting parameters and determine the proportionality factor by normalising the calculated G-V curve to the experimental value of conductance at a high bias value. The result of such fits for the two uppermost Ta/Au point contact spectra in Figure 2(a) is shown with solid lines. This analysis ignores the fact that at high bias the measured $R_d(V>>\Delta/e)$ values for these spectra contain contributions from both $R_s$ and $R_M$ whereas the spectra calculated from BTK model will have a contribution only from $R_s$. A quantititave estimate of the contribution from $R_M$ is difficult without a detailed knowledge of the I-V characteristic of the superconductor above the critical current. To get a qualitative understanding of the error involved in this kind of fits, we tried to fit the calculated curves in Figure 3(a) (generated by adding a finite contribution from $R_M$) with BTK model alone, ignoring the contribution of the finite resistivity of the superconductor above $I_c$ (see Figure 3(b)). Though with suitable choice of Z and $\Delta$ the curves can be fitted for bias voltages below and above the dips, the values of $\Delta$ are overestimated. The inset of Figure 3(a) shows how this error increases with increasing $R_M/R_s$ ratio in the spectrum. Though this procedure may introduce a small error when the dips are small, it will introduce significant error in $\Delta$ when the dips are large. Similiarly for a contact between a ferromagnet and a superconductor $P_t$ gets underestimated as the contribution from $R_M$ increases. This trend can be seen in the fits shown in Figure 2(b).

As a consistency check of the proposed explanation of the dips we can also try to estimate the critical current density ($J_c$) of the superconductor from the observed dips. When the contribution of $R_M$ in the spectra is small the normal state differential resistance is $R_d(V>>\Delta/e) \approx R_s = 2(h/e^2)/(ak_F)^2$. For Au[23] $k_F \sim 1.21 \times 10^8$ cm$^{-1}$. This gives the contact diameter $a \sim 120$Å for the Au-Ta contact with $R_d(V>>\Delta/e)=2.8\Omega$. A rough estimate of the critical current can be obtained from the voltage at which the experimental curve deviates from the BTK best fit. Comparing this voltage with the corresponding current in the I-V curve, obtained by integrating the G-V curve, we get a critical current of 0.41mA. This gives $J_c \sim 3.6 \times 10^8$ A/cm$^2$, a reasonable number considering the approximations involved.

In summary, we have presented a study of the emergence of anomalous dips in the conductance in point contacts between normal metals and conventional superconductors. From the correlation between the structure of the dips with area of contact we conclude that the dips arise from the finite resistivity of the superconducting electrode above the critical current when the contact is not in the ballistic limit. We have also shown that in the thermal limit of the point contact an enhancement of the zero bias conductance larger than twice the value at high bias can be observed if the contact is made between a good normal metal and a superconductor with large normal state resistivity. It is useful to recall that in unconventional superconductors such a kind of enhancement has been observed and often attributed to the formation of Andreev bound states. It could be worthwhile to explore the extent to which $R_M$ may contribute in the enhancement of zero bias conductance even in such systems.

*Acknowledgements:* We would like to acknowledge Profs. A. K. Nigam and E. V. Sampathkumaran for providing samples of $Au_{1-x}Fe_x$ and $Y_2PdGe_3$ respectively and Profs. S. Ramakrishnan and H. Küpfer for single crystals of $V_3Si$. We would also like to thank Professor S. Bhattacharya for encourangement and guidance and Professor A. K. Grover for critically reading the manuscript. Two of us (GS and SM) would also like to acknowledge TIFR Endowment Fund for partial financial support in the form of TIFR Alumni Association Career Development Scholarship.

**Figure Captions**

**Figure 1.** Conductance versus voltage characteristics of point contacts using different superconductors and normal metals: (a) Au foil/Nb tip at different temperatures in zero field; (b) $Au_{0.95}$-$Fe_{0.05}$ foil/Nb tip in different magnetic fields at 2.43K; (c) Fe foil/Nb tip at different temperature in zero field; (d) Au foil/Ta tip, (e) $V_3Si$ single crystal/Pt-Ir tip, (f) $Y_2PdGe_3$ polycrystalline sample/Pt-Ir tip. In (f) a zero bias enhancement of the order of 5 is observed. Sharp dips in conductance are observed for all spectra (marked by arrows) at voltage values larger than the respective superconducting energy gaps. All curves shown in (a)-(c) except the bottom most curves in each case are shifted upward for clarity.

**Figure 2.** Evolution of PCAR spectra for (a) an Au/Ta (b) Fe/Nb point contact with normal state contact resistance $R_d(V>>\Delta/e)$ of the contact (solid circles). Solid lines in the two topmost curved in panel (a) and in panel (b) are BTK fits with fitting parameters as shown in the figure. The appearance of two peaks at voltage value higher than $\Delta/e$ for $R_d(V>>\Delta/e) = 1.3\Omega$ in panel (b) is shown with arrows. Some curves have been ommited for clarity.

**Figure 3.** Current versus voltage characteristics of a superconductor/normal metal point contact generated theoretically, adding the effect of critical current over BTK model. (a) The solid and dashed lines are the I-V characteristics obtained from BTK model (with Z=0.5 and $\Delta$=0.56 meV) and the typical I-V characteristic assumed for the superconductor respectively. $R_d$ vs V curves (open circles) for different $R_M$ to $R_S$ ratio is shown in the same figure. Two sharp peaks symmetric about V=0 (shown by arrows) in $R_d$ vs. V spectra arise and become sharper with increasing $R_M/R_S$. (b) BTK fits of the curves generated in (a) neglecting the contribution of critical current. Solid lines show the fits and open circles are generated $R_d$ versus V with different $R_M/R_s$ ratio. Vertical shift has been given to all curves except the bottom one for clarity. *Inset of (a):* Variation of $\Delta$

obtained by fitting the curves generated with different $R_M/R_s$ ratio with BTK model.

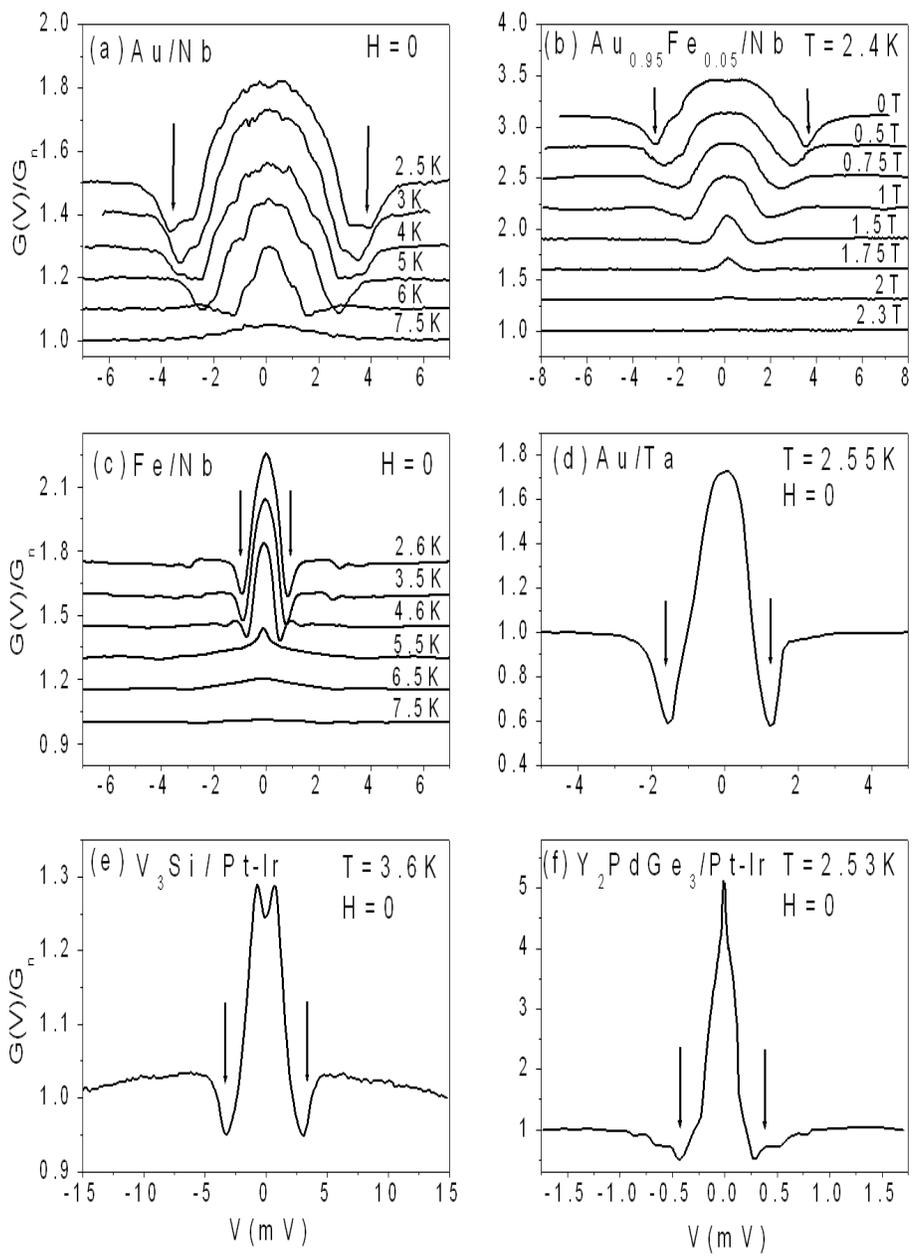

Figure 1
Author: Goutam Sheet et.al.

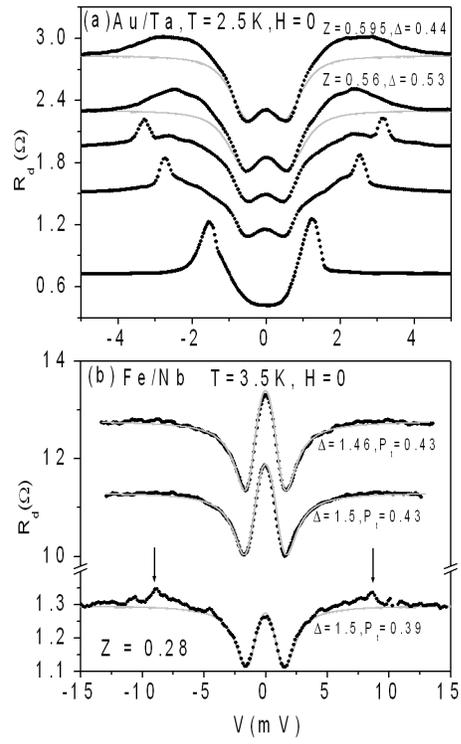

Figure 2
Author: Goutam Sheet et al.

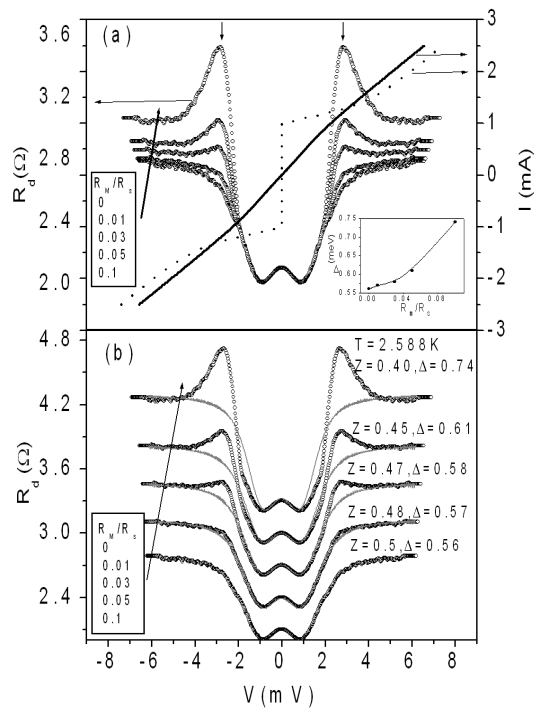

Figure 3
Author: Goutam Sheet et.al.

16